\documentclass[12pt,preprint]{aastex}

\def\swift{{\it Swift}}
\def\hete{{\it HETE-2}}

\def\prince{1}
\def\ociw{2}
\def\hubble{3}

\shorttitle{Galaxy Clusters Associated with Short GRBs II}
\shortauthors{Shin and Berger}

\begin{document}

\title{Galaxy Clusters Associated with Short GRBs. II.  Predictions
for the Rate of Short GRBs in Field and Cluster Early-Type Galaxies}

\author{
M.-S.~Shin\altaffilmark{\prince}
and E.~Berger\altaffilmark{\ociw,}\altaffilmark{\prince,}\altaffilmark{\hubble}
}

\altaffiltext{\prince}{Princeton University Observatory, Peyton Hall,
Ivy Lane, Princeton, NJ 08544}

\altaffiltext{\ociw}{Observatories of the Carnegie Institution of
Washington, 813 Santa Barbara Street, Pasadena, CA 91101}

\altaffiltext{\hubble}{Hubble Fellow}

\begin{abstract} 
We determine the relative rates of short GRBs in cluster and field
early-type galaxies as a function of the age probability distribution
of their progenitors, $P(\tau)\propto \tau^n$.  This analysis takes
advantage of the difference in the growth of stellar mass in clusters
and in the field, which arises from the combined effects of the galaxy
stellar mass function, the early-type fraction, and the dependence of
star formation history on mass and environment.  This approach
complements the use of the early- to late-type host galaxy ratio, with
the added benefit that the star formation histories of early-type
galaxies are simpler than those of late-type galaxies, and any
systematic differences between progenitors in early- and late-type
galaxies are removed.  We find that the ratio varies from $R_{\rm
cluster}/R_{\rm field}\sim 0.5$ for $n=-2$ to $\sim 3$ for $n=2$.
Current observations indicate a ratio of about $2$, corresponding to
$n\sim 0-1$.  This is similar to the value inferred from the ratio of
short GRBs in early- and late-type hosts, but it differs from the
value of $n\approx -1$ for NS binaries in the Milky Way.  We stress
that this general approach can be easily modified with improved
knowledge of the effects of environment and mass on the build-up of
stellar mass, as well as the effect of globular clusters on the short
GRB rate.  It can also be used to assess the age distribution of Type
Ia supernova progenitors.
\end{abstract}

\keywords{gamma-rays:bursts --- galaxies:clusters --- galaxies:formation}

\section{Introduction}
\label{sec:intro}

Gamma-ray bursts (GRBs) are divided into two broad classes of
short/hard and long/soft bursts \citep{kmf+93}, which appear to have
different progenitor populations.  Observations of Type Ic supernovae
in association with long GRBs provide a direct confirmation that they
arise from the death of massive stars \citep{hsm+03,smg+03}.  Short
GRBs, on the other hand, have long been suspected to arise from the
merger of neutron star and/or black hole binaries (NS-NS, NS-BH; e.g.,
\citealt{elp+89,npp92,rr02,ajm05}).  During the last year, follow-up
observations of \swift\ and \hete\ short GRBs have provided initial
confirmation to this idea, based in particular on the localization of
some short GRBs to elliptical galaxies \citep{bpc+05,gso+05,bpp+06},
their lack of association with bright supernovae
\citep{ffp+05,hwf+05,sbk+06}, and their lower energy release and wider
beaming angles compared to long GRBs \citep{bgc+06,sbk+06}.

Despite this observational progress we are still missing a clear
understanding of the progenitor population, due to the lack of direct
observations (circumburst chemical abundances, gravitational waves, or
a sub-relativistic, radioactive component: \citealt{lp98,kul05}).
Thus, statistical studies of the burst properties can be highly
effective in understanding their progenitor population(s) (e.g.,
\citealt{gno+05,ngf05,gp06}).  Recently, \citet{zr06} investigated the
ratio of short GRBs in early- and late-type galaxies as a constraint
on the age distribution of the progenitors.  Their analysis combines
the global star formation rate in each galaxy type with a local galaxy
stellar mass function, assuming the formation process of short GRB
progenitors does not depend on any other physical parameters.  Since
each galaxy type has experienced a globally different star formation
history, the ratio of bursts in each type is predicted to vary
as a function of the progenitor age distribution.

A complementary way to constrain the progenitor age distribution is to
use the rates of short GRBs in clusters and the field
(\citealt{bsm+06}; hereafter Paper I).  This approach takes advantage
of the following differences between cluster and field environment. 
First, the galaxy stellar mass
function of clusters is more heavily dominated by massive galaxies
than in the field \citep{cfn+05,bbb+06}.  Since the star formation
history is mainly determined by galaxy mass, the overall growth of
stellar mass is in turn affected by the large-scale environment.
Second, the fraction of early-type galaxies is larger in clusters than
in the field \citep{dre80,wgj93,bbb+06}.  Finally, there appears to be
a systematic offset in the star formation histories of cluster and
field early-type galaxies of $\sim 1-3$ Gyr (e.g.,
\citealt{brc+98,ksc+02,tmb+05}).  These effects lead to an overall
difference in cluster and field star formation histories, which
combined with the progenitor age distribution, is expected to affect
the relative fraction of short GRBs in each environment.

In this paper we quantify these effects and show how the ratio of
short GRBs in cluster and field early-type galaxies can be used to
understand the age distribution of the progenitors.  This is part of
our on-going systematic study of galaxy clusters hosting short GRBs,
using multi-slit optical spectroscopy and X-ray observations (Paper
I).  We find that the current observations, albeit with a small number
of events, favor $P(\tau)\propto \tau^n$ with $n\sim 0-1$.  Finally,
we provide a comparison of the systematic effects in this approach and
the approach of \citet{zr06}.

\section{The Rate of Short Bursts in Clusters and the Field}
\label{sec:theory}

The approach used by \citet{zr06} can be generalized to formulate the
short GRB rate per unit volume in cluster and field early-type
galaxies at $z\sim 0$, using the star formation history function,
${\rm SFH}(\tau)$ instead of a star formation rate function:
\begin{equation}
R_i=C\int^{t(z=\infty)}_{0} {\rm SFH}_i(\tau)\, P(\tau)\, d\tau.
\label{eqn:rate}
\end{equation}
Here $\tau$ is both the look-back time and the time delay of a short
GRB progenitor, and $i$ designates a cluster or field environment.
$P(\tau)$ represents the time delay probability distribution of the
progenitors with a normalization constant $C$.  In the context of
NS-NS or NS-BH mergers we adopt the standard power-law form, $P(\tau)
\propto\tau^n$.  We note that NS-NS binaries in the Milky Way appear
to follow $P(\tau)\propto \tau^{-1}$ \citep{clm+04}.  It is thus the
convolution of $P(\tau)$ with ${\rm SFH(\tau)}$ that determines the
relative rate of short GRBs in cluster and field early-type galaxies.

Since the star formation history of early-type galaxies is determined by
both the mass and environment of a galaxy, the total star formation 
history of each environment can be described in the following manner:
\begin{equation}
{\rm SFH}_i(\tau)=\int^{M_u}_{M_l} \phi_i(M)\, {\rm SFH}_{{\rm
gal},i}(\tau,M)\, dM,
\label{eqn:sfh}
\end{equation}
where $\phi_i(M)$ is the galaxy number density function, ${\rm
SFH}_{{\rm gal},i}(\tau,M)$ is the star formation history function of
a single galaxy that has a stellar mass, $M$, and $M_u$ and $M_l$ are
the appropriate upper and lower mass integration limits (see
\S\ref{sec:apply}).

From SDSS observations, it has been determined that the galaxy number
density is well described by a double Schechter function
\citep{bbb+06}:
\begin{equation}
\phi(M)dM = e^{-M/M^*}[\phi^*_1 (M/M^*)^{\alpha_1}+\phi^*_2
(M/M^*)^{\alpha_2}]\frac{dM}{M^*},
\label{eqn:phi}
\end{equation}
where the parameters ${\rm log}\, M^*$, $\phi^*_1$, $\alpha_1$, $\phi^*_2$ and
$\alpha_2$ for a cluster environment (${\rm log}\,\Sigma\sim 1.3$) are
11.06, 0.74, $-1.09$, 0.07, and $-1.5$, and for a field environment
(${\rm log}\,\Sigma\sim -0.9$) they are 10.44, 2.7, $-0.2$, 0.8, and
$-1.5$ \citep{bbb+06}.  Here, $\Sigma$ is the projected density of
neighboring galaxies, and the specific values are determined from a
sample of $\sim 1.5\times 10^5$ galaxies in the redshift range
$0.01-0.085$ in the SDSS Data Release Four \citep{bbb+06}.

We are here only interested in the rates of short GRBs in early-type
galaxies, whose star formation history is better understood than those
of late-type galaxies, and in order to avoid any systematic
differences between progenitors in late- and early-type host galaxies.
We therefore need to modify $\phi(M)$ by the fraction of early-type
galaxies in each environment \citep{bbb+06}:
\begin{equation}
f_{r,i}(\Sigma,M)=1-{\rm exp}\{-[(\Sigma/b_1)^{b_2}+(M/b_3)^{b_4}]\},
\label{eqn:fri}
\end{equation}
where the values of the parameters $b_1$, $b_2$, $b_3$, and $b_4$ are
$10^{0.91}$ Mpc$^{-2}$ , 0.69, $10^{10.72}$ M$_\odot$, and $0.59$.
Thus, $\phi_{i}(M)=\phi(M)f_{r,i}(\Sigma,M)$.  In a cluster
environment, with a higher $\Sigma$ and systematically larger masses,
the early-type fraction is larger than in the field.  A plot of
$f_r\phi(M)$ for cluster and field environments is shown in
Figure~\ref{fig:panels}a.  Clearly, the most massive galaxies reside
preferentially in clusters, while the bulk of the mass in galaxies
with $M\lesssim 10^{10.5}$ is in the field.

Finally, we turn to the star formation history function, ${\rm
SFH_{gal}}(\tau,M)$.  Early-type galaxies in different environments
show different star formation histories for a given galaxy mass
\citep{ksc+02,tmb+05,dsw+06,schawinski06}.  Moreover, the star formation history is
also determined by the galaxy mass.  Following \citet{tmb+05}, we use
a Gaussian form for the star formation history:
\begin{equation}
{\rm SFH_{gal}}(\tau,M)=\frac{M}{\sqrt{2\pi}\Delta t}{\rm exp}
[-\frac{(\tau-t_{\rm peak})^2}{2(\Delta t)^2}],
\label{eqn:sfhgal}
\end{equation}
where the peak of the star formation history function, $t_{\rm peak}$,
is determined by both the mass of a galaxy and its environment
(Equations 2 and 3 of \citealt{tmb+05}), and $\Delta t$ is determined
by the mass of a galaxy.  Here, we assume that the low-density
environment of \citet{tmb+05} corresponds to the field, while clusters
correspond to the high-density
environment\footnotemark\footnotetext{Following \citet{tmb+05} we
adopt the following cosmological parameters: $\Omega_m=0.3$,
$\Omega_\Lambda=0.7$, and $H_0=75$ km s$^{-1}$ Mpc$^{-1}$.}.  The
overall trend is that less massive galaxies form their stars later 
(i.e., smaller $t_{\rm peak}$) and over a wider timescale (i.e.,
larger $\Delta t$).  In addition, $t_{\rm peak}$ in clusters is
systematically larger than in the field.  These effects are shown in
Figure~\ref{fig:panels}b, where we plot the star formation histories
of galaxies with $10^8$, $10^9$, $10^{10}$, and $10^{11}$ M$_\odot$ in
both environments.

The final ingredient is the overall normalization of
Equation~\ref{eqn:phi}.  In its current form this equation was
normalized to a total mass of $10^{10}$ M$_\odot$ for each environment
by \citet{bbb+06}.  Since we are interested in the overall rate per
unit volume, we therefore need to know what fraction of the stellar
mass is in clusters versus the field at $z\sim 0$.  From the study of
\citet{bbb+06}, as well as \citet{fhp98} and \citet{ebc+05}, it
appears that about $20\%$ of the total stellar mass is included in
cluster environments\footnotemark\footnotetext{This corresponds to the
assumption that cluster environments can be described by ${\rm log}\,
\Sigma\gtrsim 0.5$; see Appendix A of \citet{bbb+06} for details.},
which we adopt here.

\section{Constraints on the Age Distribution of Short GRB Progenitors}
\label{sec:apply}

To illustrate the combined effect of the trends discussed in the
previous section we begin by making the simplified assumption that
there is a single typical galaxy mass, $M_{\rm typ}$, for each
environment.  This typical mass in turn determines the typical star
formation history of each environment.  In this scenario,
Equation~\ref{eqn:sfh} can be simplified as:
\begin{equation}
{\rm SFH}_i(\tau)=\phi_i(M_{\rm typ})\times{\rm SFH}_{{\rm
gal},i}(\tau,M_{\rm typ}),
\label{eqn:sfhsimple}
\end{equation}
while $M_{\rm typ}$ is determined by \citep{bbb+06}:
\begin{equation}
{\rm log}\,(M_{\rm typ})=10.73+0.15{\rm log}\,\Sigma.
\end{equation}
Therefore, $M_{\rm typ}$ is $\sim 10^{10.6}$ M$_\odot$ and $\sim
10^{10.9}$ M$_\odot$ for field and cluster environments, respectively.

\begin{figure}
\plotone{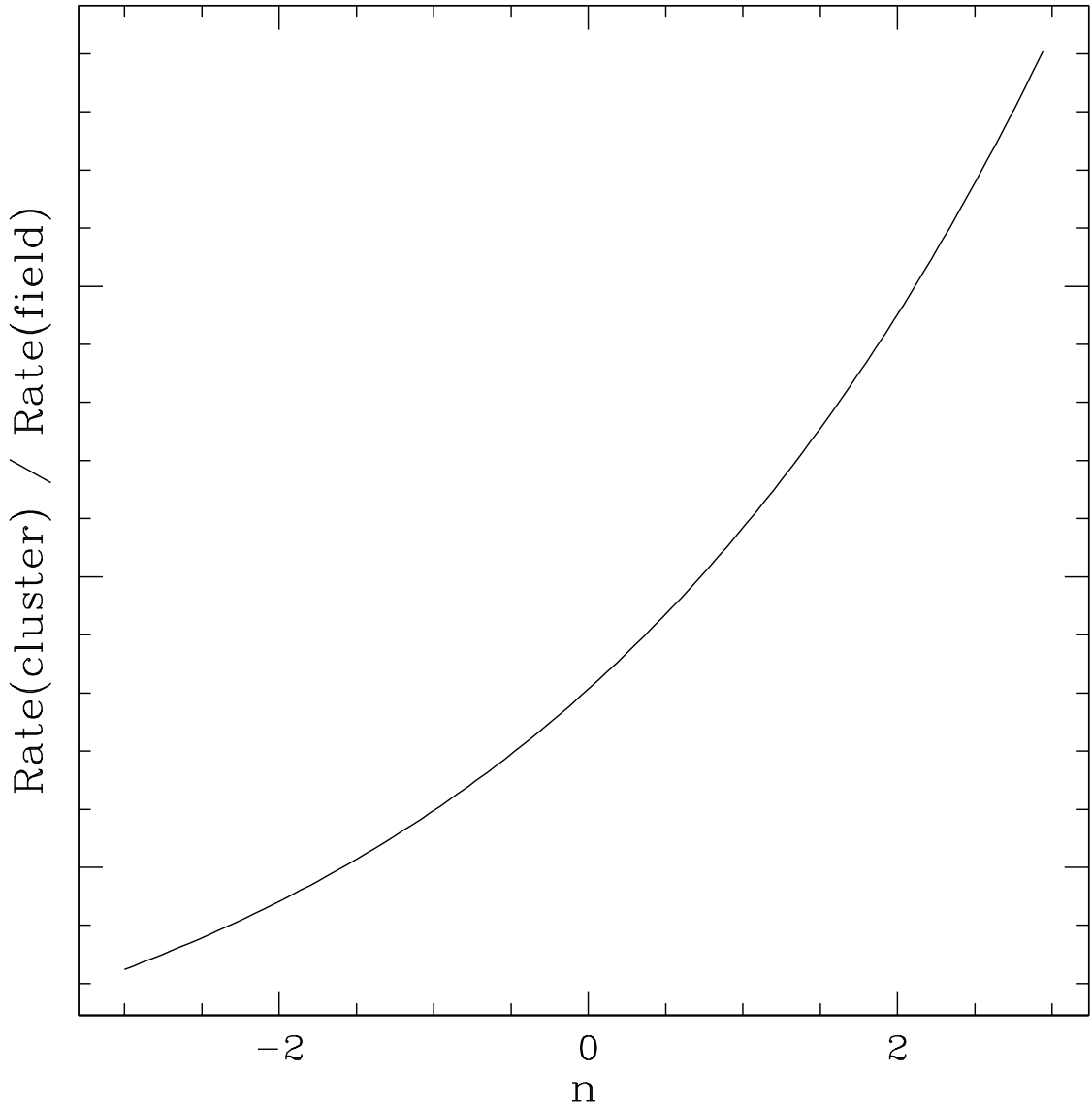}
\caption{Ratio of the short GRB rate in cluster and field early-type
galaxies as a function of the age distribution power law index, $n$,
assuming that each environment is described by a typical galaxy mass
(\S\ref{sec:apply}).  The scale on the ordinate is arbitrary since we
do not consider the full mass function, but the overall trend is
representative.  Since the typical galaxy mass is higher in clusters
than in the field, the typical star formation epoch is earlier in
clusters.  We therefore expect more short GRBs in cluster early-type
galaxies when $n$ is high.
\label{fig:simple}}
\end{figure}

We are now in a position to use Equations~\ref{eqn:rate},
\ref{eqn:phi}, \ref{eqn:fri}, \ref{eqn:sfhgal}, and
\ref{eqn:sfhsimple} to predict the relative rate of short GRBs in
cluster and field early-type galaxies.  The result as a function of
the power law index $n$ is shown in Figure~\ref{fig:simple}.  We note
that the absolute scale in this plot is irrelevant since we did not
integrate over the full mass function, but it provides insight into
the overall trend.  Namely, as $n$ increases, i.e., as the
distribution is more heavily weighted to older progenitors, the
fraction of short GRBs in clusters increases.  This can be understood
as the combined effect of a systematically earlier star formation
episode and a higher typical mass in clusters.

\begin{figure}
\plotone{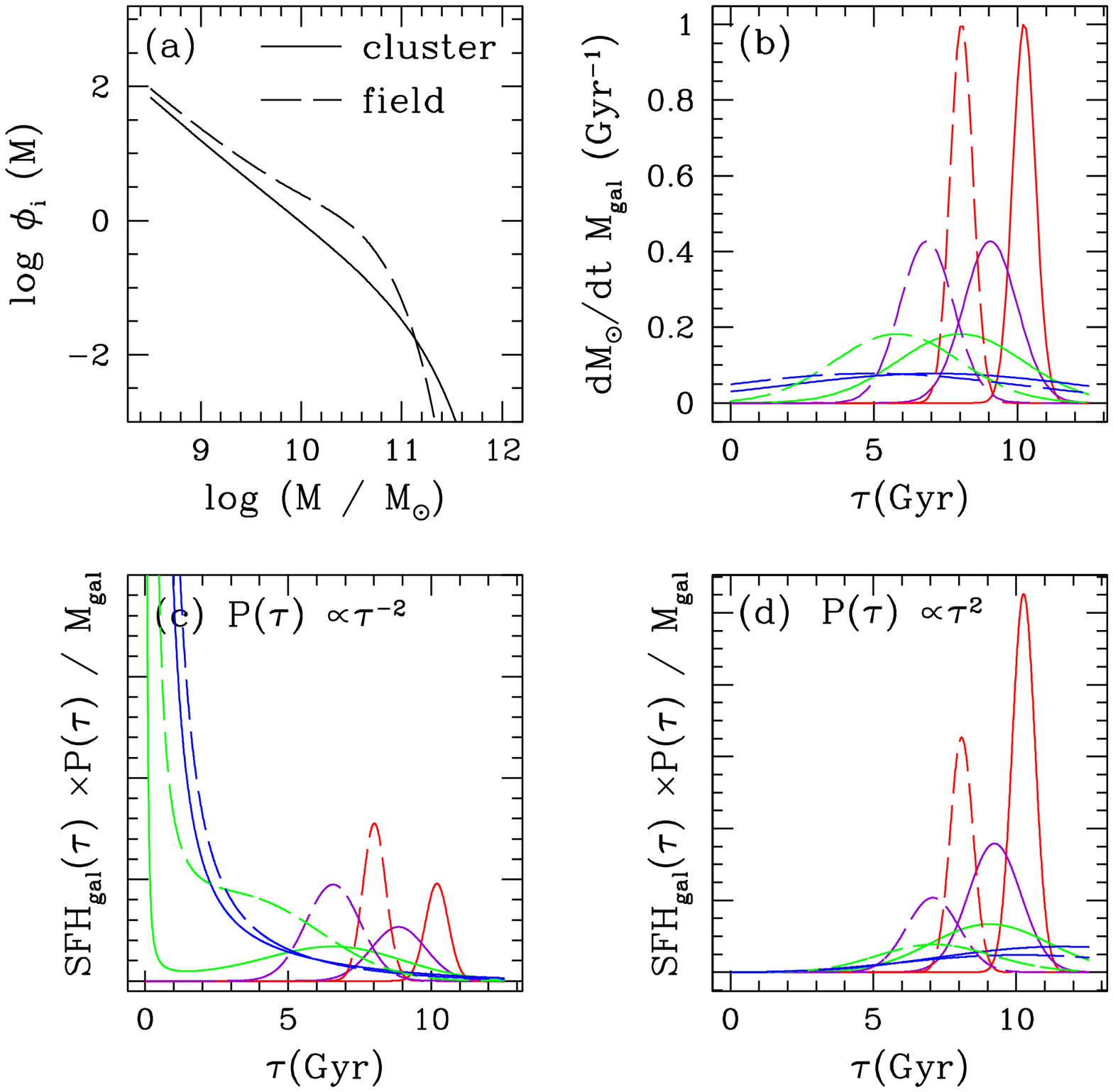}
\caption{A summary of the model ingredients that determine the
relative rate of short GRBs in cluster and field early-type galaxies.
Panel (a) shows the mass function in each environment (solid line:
cluster; dashed line: field), including the different early-type
fractions (Equation~\ref{eqn:fri}), and an overall cluster mass
fraction of $20\%$ (\S\ref{sec:theory}).  Panel (b) shows the star
formation history as a function of environment and galaxy mass (red:
$10^{11}$ M$_\odot$; purple: $10^{10}$ M$_\odot$; green: $10^9$
M$_\odot$; blue: $10^8$ M$_\odot$).  Bottom panels show the product of
the star formation history with the short GRB progenitor age
distribution function for a power law index $n=-2$ (c), and $n=2$ (d).
Clearly, a distribution weighted to short merger timescales heavily
favors lower mass host galaxies (and hence the field), while a
distribution weighted to long merger timescales favors massive host
galaxies (and hence clusters).
\label{fig:panels}}
\end{figure}

A quantitative determination of the relative rates requires a full
integration of Equation~\ref{eqn:sfh}.  The results of this
integration are shown in Figure~\ref{fig:full}.  Overall, the
dependence of the ratio of short GRBs in clusters and the field on $n$
is similar to the one found in the simple case.  However, the
integration over the full mass function of each environment brings out
additional trends.  For a low mass cutoff smaller than $10^{10}$
M$_\odot$, there is sharp downturn in the ratio for $n\lesssim -1$.
This can be understood from the plots of ${\rm SFH}_{\rm
gal}(\tau)\times P(\tau)$ shown in Figures~\ref{fig:panels}c and
\ref{fig:panels}d.  In particular, for $n=-2$ the short GRB rate in
the field is larger than in clusters since the field has a higher
abundance of low mass ($\lesssim 10^{10}$ M$_\odot$) early-type
galaxies (Figure~\ref{fig:panels}a), which have young or intermediate
stellar populations.  For such low values of $n$ the ratio is
sensitive primarily to recent star formation.  We note that the same
effect is seen in the analysis of \citet{zr06} for the ratio of early-
to late-type hosts.

In this context ($n\lesssim -1$) the lower mass cutoff in
Equation~\ref{eqn:sfh} plays an important role.  In
Figure~\ref{fig:full} we show the effect of setting $M_l=10^8$,
$10^9$, and $10^{10}$ M$_\odot$.  As noted above, for $M_l=10^{10}$
M$_\odot$ we do not see an obvious downturn because galaxies above
this limit do not exhibit any obvious recent star formation activity.
For the lower values of $M_l$, we find that the largest downturn is
for $M_l=10^{9}$ M$_\odot$.  The reason for this is evident in
Figure~\ref{fig:panels}a, which shows that the largest difference
between the field and cluster mass functions is at $M\gtrsim 10^9$
M$_\odot$.  At lower masses the mass functions converge, leading to a
ratio that is $\sim 1$.  

For $n\gtrsim -1$, on the other hand, the rate is dominated by the
oldest, and hence most massive galaxies (Figure~\ref{fig:panels}d).
In this case the ratio does not depend on $M_l$, and the predominance
of massive galaxies in clusters, along with the systematically earlier
star formation episodes, results in an increased fraction of short
GRBs in clusters for larger $n$.  In fact, for $n=2$, we find that
there are three times as many short GRBs in cluster early-type
galaxies as there are in field early-type galaxies.

\section{Discussion}
\label{sec:disc}

\begin{figure}
\plotone{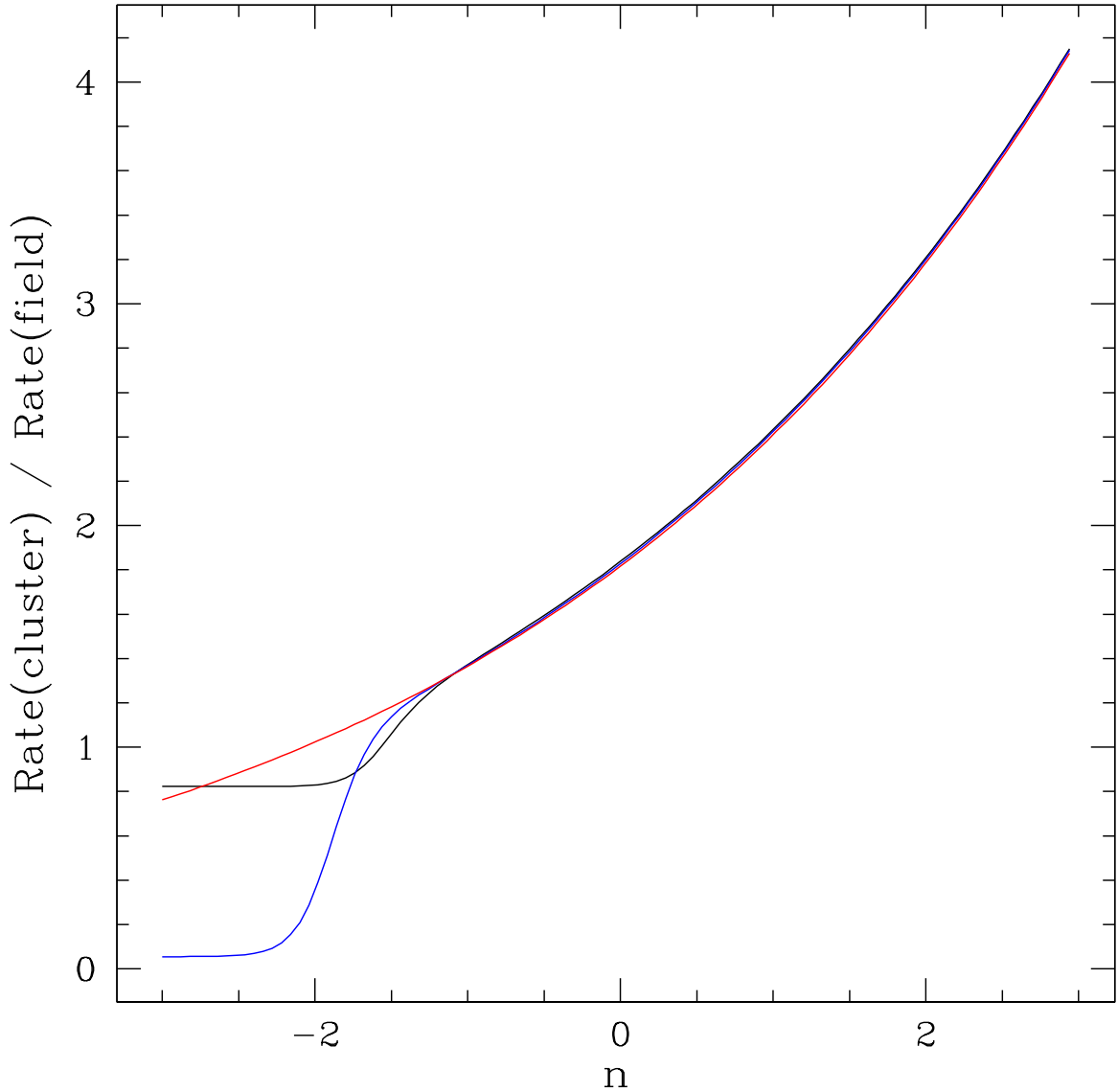}
\caption{Ratio of the short GRB rate in cluster and field early-type
galaxies as a function of the age distribution power law index, $n$,
considering the full mass function in each environment.  Different
colors represent the effect of different mass integration limits:
$10^8<M<10^{12}$ M$_\odot$ (black), $10^9<M<10^{12}$ M$_\odot$ (blue),
and $10^{10}<M<10^{12}$ M$_\odot$ (red).  The current observed ratio,
based on only three events, is about 2, suggesting that $n\sim 0-1$.
\label{fig:full}}
\end{figure}

We now turn to a comparison of our model with observation of short
GRBs.  To date two short bursts have been localized to clusters at
$z\sim 0.2$: GRB\,050509b \citep{gso+05,bpp+06} and GRB\,050911
(Paper1).  We note that in the latter case the large error
circle prevents an association with a specific cluster galaxy, but the
large early-type fraction of $80\%$ \citep{bsm+06} suggests that the
burst was likely hosted by an early-type galaxy.  On the other hand,
only one short GRB has been localized to a field early-type galaxy,
GRB\,050724 \citep{bpc+05}.  We do not consider GRB\,050813, which was
hosted by a cluster at a much higher redshift ($z\sim 1.8$;
\citealt{ber06}), and GRB\,060502b, which may be hosted by an
early-type galaxy \citep{bpc+06}, but whose large-scale environment
has not been fully explored yet\footnotemark\footnotetext{The limit on
the X-ray luminosity at the redshift of the putative host galaxy,
$L_X\lesssim 6\times 10^{42}$ erg s$^{-1}$, is lower by a factor of
eight than that of the cluster hosting GRB\,050509b, but it is
somewhat higher than that of the cluster hosting GRB\,050911
(Paper I).}.  Thus, the current ratio of short GRBs in cluster
versus field early-type galaxies is about $2\!:\!1$, with a large
uncertainty due to the small number of events.  From
Figure~\ref{fig:full}, we find that this ratio corresponds to $n\sim
0-1$. This value is lower than $n\gtrsim 3/2$ claimed by \citet{zr06}, 
but is in rough agreement with $-1 \lesssim n \lesssim 0$ found by \citet{bfp+06} based 
on their revised redshift distribution with $1/4-2/3$ of all short GRBs at 
$z\gtrsim 1$.

Clearly, in both methods of estimating the age of short GRB
progenitors the uncertainty in the inferred value of $n$ is currently
dominated by the small number of bursts with a known redshift, host
galaxy, and large-scale environment type.  Since this uncertainty will
eventually diminish with a larger sample of events, it is interesting
to consider systematic uncertainties in both theoretical approaches.
Our analysis suffers from the somewhat poor definition of cluster and
field environments.  We have used an overall cluster mass fraction of
$20\%$, as indicated by several researchers, but this number may range
from $10$ to $30\%$.  Second, we have used the simplified bimodal star
formation history model of \citet{tmb+05}, but these authors do not
use the same quantitative definition of galaxy environment that was
used for the mass functions by \citet{bbb+06}.  Since we have used
representative mass functions, and then scaled the results by the
overall mass fraction in clusters and the field, this effect should
not be significant. Third, the uncertainty in the definition of field 
galaxy environment leads to an overall uncertainty of about $20\%$ in 
our calculated ratio. Finally, since for $n\lesssim -1$ the ratio
depends on $M_l$, it is essential to understand the appropriate low
mass limit for early-type galaxies.  With the current inferred value
of $n\sim 0-1$, however, this may not be a relevant issue for short
GRBs.

Similarly, the analysis of the short GRB rate in early- and late-type
galaxies performed by \citet{zr06} also suffers from systematic
effects.  First, the ratio is affected by the uncertain star formation
history at high redshift when early-type galaxies formed most of their
stars.  Second, their estimation of star formation rate does not
consider environmental effects that appear to be important in current
observations from SDSS and 2dF, and which we have accounted for here.
Finally, as noted by \citet{zr06}, it is possible that late-type
galaxies have an altogether different age distribution of short GRB
progenitors than early-type galaxies.  This problem is overcome by our
method since it considers only early-type host galaxies.  We note that
if both derivations are in fact correct, then the estimated values of
$n$ can be used to assess any systematic differences of short GRB
progenitors in early- and late-type galaxies, as suspected to exist
for type Ia supernovae \citep{slp+06}.

Future applications of our approach will include the effect of
globular clusters, which are thought to provide an efficient
environment for the production of NS-NS binaries, and may account for
a substantial fraction of all short GRB progenitors \citep{gpm06,hopman+06}.  We
expect that since the specific frequency of globular clusters
increases significantly with galaxy mass \citep{har91}, an association
with globular clusters will increase the fraction of short GRBs in
galaxy clusters compared to the trend shown in Figure~\ref{fig:full}.
Similarly, our approach can be extended to higher redshift to
investigate the evolution in the fraction of short GRBs in clusters.
We expect that the lack of strong evolution in the last several Gyr 
likely makes our analysis applicable out to $z\sim 1$.  However, if
some short GRBs are in fact associated with clusters at $z\sim 2$,
this presents an opportunity to assess any systematic changes in the
value of $n$ with redshift.

We end with the following conclusion.  If our current estimate of
$n\gtrsim 0$ continues to be supported by future observations, then
this implies that the majority of short GRBs in early-type galaxies
will occur in clusters.  This therefore suggests that short GRBs can
provide an efficient tool for finding forming galaxy clusters at high
redshift, as already appears to be the case for GRB\,050813
\citep{ber06}.  Continued near-IR imaging and optical spectroscopic
observations of short GRB fields may therefore provide an efficient
method for finding the highest redshift clusters.

\acknowledgements 
We thank I.~K.~Baldry, Y.-T.~Lin, M.~Brown,
A.~Dressler, and J.~Mulchaey for helpful discussions.
M.-S.~S.~acknowledges support from the Observatories of the Carnegie
Institution of Washington. 
E.B.~is supported by NASA through Hubble Fellowship grant
HST-01171.01 awarded by STSCI, which is operated by AURA, Inc., for
NASA under contract NAS5-26555.

\end{document}